\documentstyle[aps,graphicx,prl,twocolumn]{revtex}
\tighten 
\begin{document}
\draft
\twocolumn[\hsize\textwidth\columnwidth\hsize\csname 
@twocolumnfalse\endcsname 
\title{  Dimensional Crossover of Localisation and Delocalisation in a Quantum
  Hall Bar }
\author{ Stefan Kettemann}
\address{ I. Institut f\" ur Theoretische Physik, Universit\" at Hamburg,
  Jungiusstra\ss{}e 9, 20355  Hamburg, Germany}
\maketitle

\begin{abstract}

 The  2-- to 1--dimensional crossover 
  of the  localisation length 
 of electrons confined to a disordered quantum wire of 
 finite width $L_y$
 is studied 
  in a model of electrons  moving in the potential of  
 uncorrelated impurities.
  An   analytical formula for the localisation length is derived,
 describing  the dimensional crossover 
 as function of width $L_y$, conductance $g$ and  perpendicular 
  magnetic field $B$ . 
 On the basis of these results, 
 the scaling analysis of the quantum Hall effect 
 in high Landau levels, and the delocalisation 
 transition in a quantum Hall wire are reconsidered. 
\end{abstract}
\vskip2pc]

\section{INTRODUCTION}
\label{sec:introduc}
 The 
 Hall conductance of  a 2--dimensional electron system in a strong
 magnetic 
 field \cite{klitzing}  
 is precisely quantised  due to the trapping  of  electrons to
 localized states in the bulk of the
 system.  Thereby,   a change of electron density does not result in a 
  change of the 
  Hall conductance\cite{laughlin,aoki,halperin}.
 In the tail of the Landau bands
 the localisation length $\xi$  is small, on the order of 
 the cyclotron   length $l_{cyc} = \sqrt{2n +1} l_B$, where 
 the magnetic length $l_B$ is defined by $l_B^2 = \hbar/(q B) $. 
 The localisation length increases towards the middle of the Landau bands,
 located at energies 
 $E_n = \hbar \omega_c (n +1/2) $, where  $\omega_c = q B/m$ is the cyclotron frequency, 
 q being the electron charge, and $n=0,1,2,...$.
  In an infinite  system  the localisation
 length  of an  Eigen state with energy $E$  is expected to 
 diverge   as   $ \xi \sim (E -E_n)^{-\nu}$.  
 The exponent $\nu$ is found  for
 the lowest two Landau bands, $n=0,1$ to be 
   $\nu \approx 2.3$ for spin split Landau levels,
 as supported by analytical\cite{milnikov,meraikh}, 
 numerical \cite{hucke,huckerev} and experimental studies \cite{scalingexp},
 reminiscent of a second order transition from an insulator  to a metal.  
  Thus, for a finite system, there should exist 
  in the middle of a disorder broadened Landau--band, $E_n$, n being the Landau index,    a  band
 of states, which extend through the whole system of size  $L$, 
with band width 
 $\Delta E = (l_{cyc}/L)^{1/\nu} \Gamma$, where $\Gamma =\hbar \sqrt{2/\pi} \sqrt{\omega_c/\tau}$.  

 On the other hand, the localisation length in two--dimensinal systems 
 with broken time reversal symmetry is  from  the 
 one parameter scaling theory  expected  to depend exponentially 
 on the conductance $g$ as \cite{ab,weg,hikami,elk} 
\begin{equation} \label{2db}
\xi   \sim  \exp ( \pi^2 g^2 ).
\end{equation} 
$g$ is   the conductance parameter per spin channel.  
 $g$ exhibits the Shubnikov-de-Haas oscillations
 as function of the magnetic field, 
 for $\omega_c > 1/\tau$, where $1/\tau$ is the elastic scattering rate. 
  The  maxima occur,  when the
 Fermi energy is in the middle of the Landau band. Thus, 
 the localisation length is expected to increase strongly
 from the tails
 to  the middle of the Landau bands, irrespective  of  the existence of the 
quantum critical  point. 
 For uncorrelated impurities,  within  self consistent Born  approximation\cite{ando},  
 one finds  that  the maxima in the longitudinal conductance  are given by 
 $ g ( E = E_n)  = \frac{1}{\pi} (2 n + 1) = g_n $.
   Thus, one gets     localisation lengths 
 $ \xi_{2 D}( E_n) = l_{\rm cyc} \exp ( \pi^2 g_n^2 ) $  in the middle of  higher  Landau levels, $n> 1 $, 
 which are  macroscopically large\cite{huckerev,levitation}. 
However,   when the width of the wire $L_y$ is smaller than the length scale  $\xi$, 
 the localisation is expected to become quasi--1--dimensional. That is, the electrons in  the 
 middle of a  Landau band can diffuse freely between the edges of the wire,
   but are localised along the wire.
 The quasi-1-dimensional localisation length is known to depend
 only linearly on the conductance, 
 and is,  in a magnetic field, with  broken  time reversal symmetry, 
  given by\cite{ef,larkin,dorokhov,prbr}
\begin{equation} \label{1db}
\xi = 2 g (B)  L_y.  
\end{equation}
  Thus, for $ L_y \ll  \xi_{2D} (E_n)$ there is  a crossover
 from 2-- to 1--dimensional localisation as the 
  Fermi energy is moved from the tails  into the middle of
 a  Landau band.
 It is known from numerical \cite{hucke} and analytical 
 studies \cite{meraikh}, that in a finite quantum Hall bar, the criticality
  in the middle of the Landau band
 results in a finite localisation length $\xi_{crit} \approx 1.2 L_y$, 
 exceeding the wire width $L_y$.
 Comparing this value with the quasi--1--dimensional localisation length 
 for uncorrelated impurities in the middle of the Landau band, 
 obtained from Eq. (\ref{1db}), 
 $\xi_{n 1 D} =\frac{2}{\pi} (2 n+ 1) L_y$,  we see that this  length scale 
 exceeds the {\it critical} localisation length   $\xi_{crit}$ in all but the lowest  Landau level, 
 $n = 0$. 

 It is the aim of this article to derive analytically the 
 dimensional crossover of the localisation length $\xi$ in a   wire as
 function of a perpendicular magnetic field.
 This might also help to identify the irrelavent scaling parameters observed in numerical 
 studies of the integer quantum Hall transition\cite{hucke,huckerev}.   
 We also  find  that in quantum Hall bars
 of  finite width $L_y$, 
  there  exists  at low temperatures a new phase, when 
  the phase coherence length
 $ L_{\varphi}$
 exceeds  the  quasi--1--dimensional localisation length,
 in the middle of the Landau band,  
 $ L_{\varphi} > \xi_n$\cite{mesoqh}. This phase may
    accordingly be called     the mesoscopic quantum Hall phase, 
 exhibiting plateaus in the Hall conductance,
 when the bulk localisation length is smaller than the wire width, 
 and the Hall conductance is carried by edge states, 
 separated by regions in energy 
 where all states are localised along the wire, and the conductance is zero.  A similar result      has  also been noted recently 
 in a renormalization group study of a quantum Hall 
 bar in Ref. \cite{meraikh}.


  In the next section the crossover between 1--dimensional 
 and 2--dimensional localisation is studied. 
 In the third section the 
 localisation length is derived   as function of  magnetic field.  
  In the fifth section, the scaling theory of the integer quantum Hall 
 effect is considered. It is shown  that the irrelevant scaling 
 parameter and the  scaling function is in qualitative agreement 
 with the one   obtained 
 from the noncritical theory.  
 In the last section, a summary is given,
 and  implications of these results for the theory of 
 the integer quantum Hall effect are pointed out. 
 The derivations are given in appendices 
 A ( orthogonal localisation length as function of wire width $L_y$), 
 B ( unitary  localisation as function of wire width $L_y$), 
 and C (orthogonal to unitary crossover of 2D localisation length). 
 In Appendix D a generalized derivation of the field theory 
 including an edge state potential and the topological term in 
 the presence of a magnetic field is given. 

\section{ Dimensional crossover of localisation }

 In the follwing,  we study  the 
   localisation length  $\xi$  of electrons confined to   a
  disordered  wire of finite 
 width $L_y$.


{\it Orthogonal Regime. }  First, let us consider the problem 
 without magnetic field, $B=0$, the orthogonal regime.

 An estimate of  the localisation length $\xi$   can be obtained 
 by performing a  perturbative renormalization of the dimensionless conductance
  $g$, which appears as the coupling constant of 
  the  action in   the nonperturbative theory of disordered electrons.
  The bare conductance   $g$ is obtained  in self consistent Born approximation\cite{ando}. 
  The vanishing of the renormalised conductance $\tilde{g} \rightarrow 0$ as one increases the wavelength 
 of renormalisation,  signals the localisation, and can be used to obtain an estimate of the 
 localisation length $\xi$, as done in  appendix A. 

 The  first  order in $1/g$ of the  perturbative renormalization 
  corresponds to
  the weak localisation correction to the conductivity.
  Thus,  one can estimate  the localisation length
 $\xi$ at zero magnetic field $B=0$ as done in appendix A.  
 There,  the renormalisation is performed for arbitrary finite widths $L_y$
 of the wire. 
  One thus gets the   following equation for the localisation length $\xi$, 
\begin{equation} \label{lcgeneral}
\xi =  g L_y - \frac{L_y}{\pi} \ln \left[ 2 \frac{k_0 L_y/(2
    \pi)}{1 + \sqrt{ 1+ (L_y/(2 \pi \xi))^2} } \right].
\end{equation}
 where $k_0 = 2\pi/l= \pi k_F/g$, 
 and we assumed  that the wire width is diffusive,  $L_y > l$. 

 In the quasi-1-D limit, $\xi  \gg L_y$, we find a  logarithmic correction,
 to the expected quasi-1- D result, $ \xi = g L_y$: 
\begin{equation} \label{q1d}
\xi =   g L_y  - \frac{L_y}{\pi} \ln (k_F L_y/g).
\end{equation}

 In the opposite limit,  $\xi  \ll L_y$, the 
 nonlinear equation for the localisation length simplifies to, 
\begin{equation}
\xi = l \exp (  \pi g ) \exp (- \pi \frac{\xi}{L_y}).
\end{equation}

\begin{figure}
\begin{center}
\vspace{-1cm}
\includegraphics[width=.54 \textwidth]{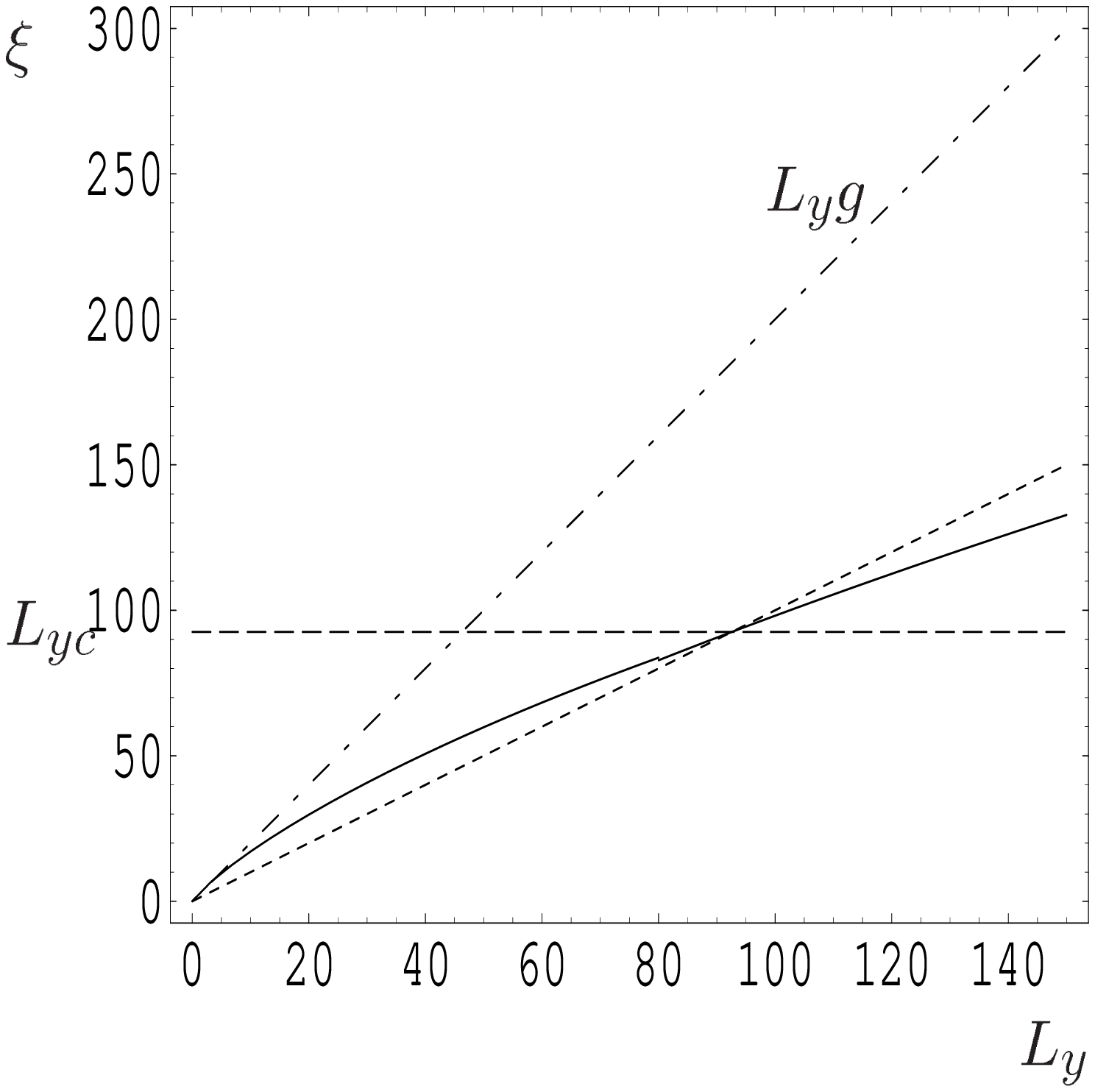}
\vspace{-6cm}
\caption{ The width  dependence of the localisation length $\xi$ 
 at fixed conductance $g=2$  without magnetic field, $B=0$, is shown as the full line.
 For comparison, the quasi--1--dimensional 
limit $ L_y g$ is drawn  as the dash--dotted line. The crossover 
 occurs  at   $\xi = L_{y c}  = \exp ( -\pi  ) (2 g/k_{\mbox{\tiny F}}) \exp (\pi g )   $. 
 }
 \label{loclengthW}
\end{center}
\end{figure}
 The solution of this equation can be written in closed form 
 in terms of the  Lambert--W--function $W_0(z)$\cite{lambert}:
\begin{equation} \label{lclambert}
\xi = \frac{L_y}{\pi} W_0 ( \frac{2 \pi}{L_y} \frac{g}{ k_{\mbox{\tiny F}}} \exp (\pi g)), 
\end{equation}
 where we substituted $ l = 2 g/ k_{\mbox{\tiny F}}$. 
 The Lambert--W--function $W_0 (x)$  
 is defined as  the solution of the nonlinear equation\cite{lambert}
\begin{equation}
z = a \exp ( - b z ),
\end{equation}
  given by 
\begin{equation}
 z = \frac{1}{b} W_0(ab). 
\end{equation}

 Thus, the localisation length is found to increase linearly with the width 
 like  
\begin{equation} \label{1d}
\xi \mid_{\xi \gg L_y}  =  g  L_y,
\end{equation}  
 when the localisation length exceeds  the width $L_y$, 
 $\xi  \gg L_y$. 
 It  logarithmically  deviates  from this behavior  
   when the width $L_y$ is on the order of  
 $ L_{y c} = \exp ( -\pi  ) l \exp ( \pi g) $.
  For larger widths,  it slowly  saturates towards 
 the width independent 2D--localisation length, 
\begin{equation} \label{2d}
 \xi \mid_{\xi \ll L_y} =  \frac{2 g}{ k_{\mbox{\tiny F}}}  \exp (  \pi g ),
\end{equation}
 as seen in Fig. \ref{loclengthW}. 

 For internediate widths there is a wide  regime where the localisation lengths 
 deviate strongly from both 1--D, Eq. (\ref{1d}),  and 2--D, Eq. (\ref{2d}),
 which would yield $\xi_{2D} ( g= 2) = 2142$ on the scale $1/k_F$, used in Fig. \ref{loclengthW}.

\begin{figure} 
\begin{center}
\includegraphics[width=.54 \textwidth]{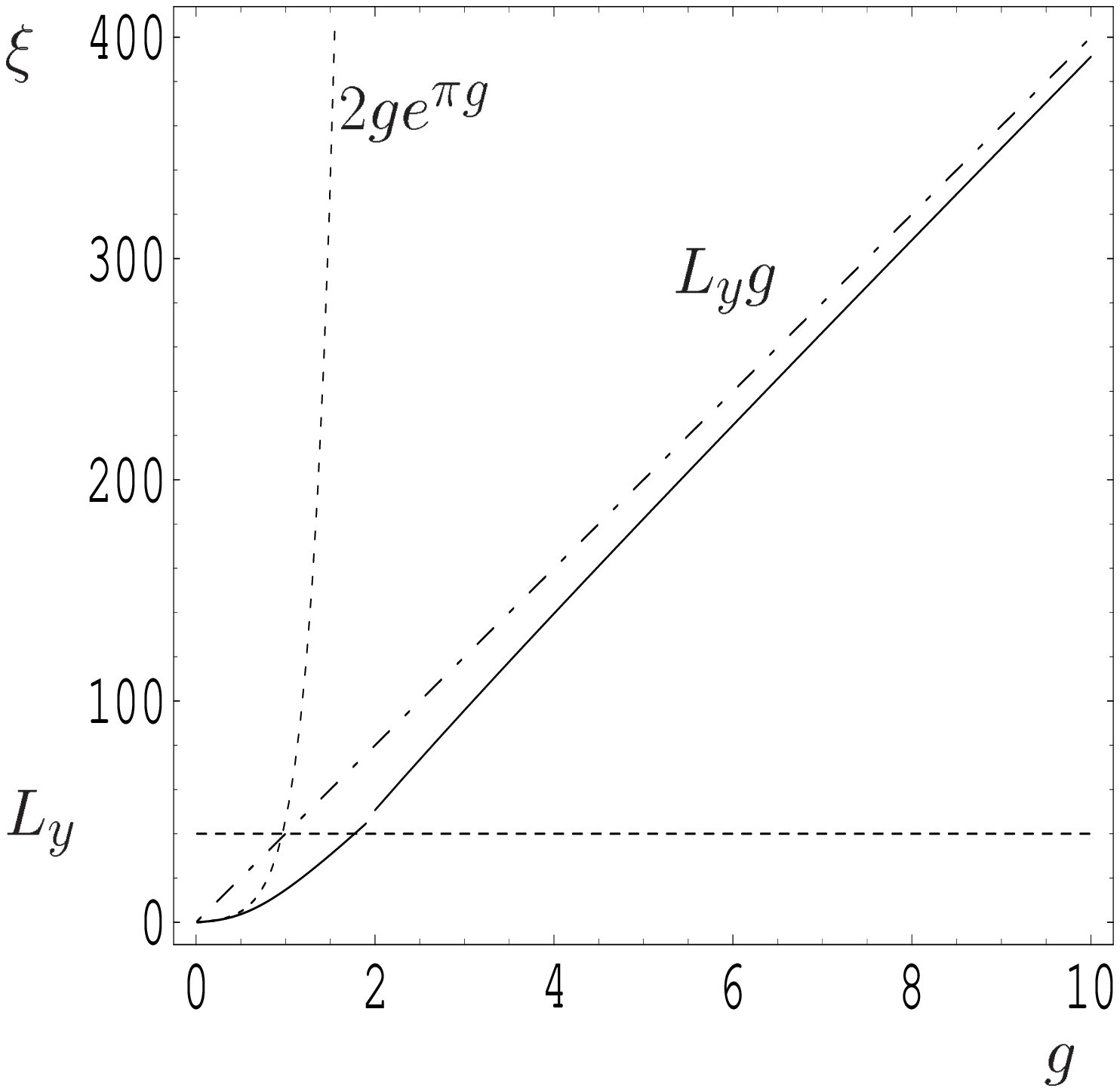}
\vspace{-6.4cm}
\caption{ The  localisation length $\xi$ (fulll line) in units of $k_{\mbox{\tiny F}}^{-1}$
  as function of  the   conductance $g$ at fixed width $L_y =40 $,
 and  without magnetic field, $B=0$. 
It is plotted using  Eq. (\ref{lclambert}) for $\xi < L_y$, and Eq. (\ref{q1d}) for $\xi > L_y$. 
 For comparison, we show the quasi--2--dimensional $\xi_{2D} = (2 g/ k_{\mbox{\tiny F}}) \exp ( \pi g )$
 (dashed line) 
 and the  quasi--1--dimensional
  $ L_y g$ (dash--dotted line) limiting functions. 
 }
 \label{lcg}
\end{center}
\end{figure}
 As one fixes the width $L_y$,  and increases the dimensionless conductance $g$, 
 a crossover from 2D-- to 1D--localisation is observed,  as shown in Fig. \ref{lcg}. 
 There,  Eq. (\ref{lclambert})
 is plotted as function of $g$, for $\xi < L_y$. For $\xi > L_y$, 
 the localisation length is plotted using, Eq. (\ref{q1d}). 
 In the intermediate regime, the solution of the general equation Eq. (\ref{lcgeneral})
 deviates only little  from  these  asymptotic solutions.  
  Note that the validity of the derivation is  limited to  $g > 1$, 
  while  the deviations from 2D localisation behavior
  occur for the chosen  width $L_y$  already close to  $g=1$. 
 The derivation given above has been done for wires of diffusive width,
 $l < L_y$, or $g < L_y k_{\mbox{\tiny F}}/2$, corresponding to $g <  20$ in Fig. 2. 

{\it Unitary Regime.}
 At moderately strong magnetic field,
 the time reversal symmetry is broken, and the so called unitary 
 regime is reached, 
 when the
 localisation length  exceeds the magnetic diffusion length\cite{mazza}, $\xi > L_B $, 
 (where $L_B = l_B$,  when $ l_B < L_y$ and $ L_B = (3)^{1/2} l_B^2/L_y$, when $l_B > L_y > l$).  
 Then,  
 the first order,  weak localisation correction vanishes and one needs to do the perturbative renormalisation to
 second  order in $1/g$. 
 Thus, we have  to calculate all diagrams contributing to this order, 
 the so called Hikami boxes\cite{hikami}, to study the dimensional crossover in a magnetic field. 
 An efficient way to do this, is to start from the supersymmetric 
 nonlinear sigma model and do an expansion around its 
 classical point, 
 as done in  Ref. \cite{ef}
 for the pure orthogonal and unitary regimes  in 2 dimensions.

  Performing  this  renormalisation for wires of finite width $L_y$ in the unitary regime, we obtain
 in appendix B,  that the 
 localisation length $\xi$  satisfies in the unitary regime the equation, 
\begin{figure} 
\begin{center}
\includegraphics[width=.54 \textwidth]{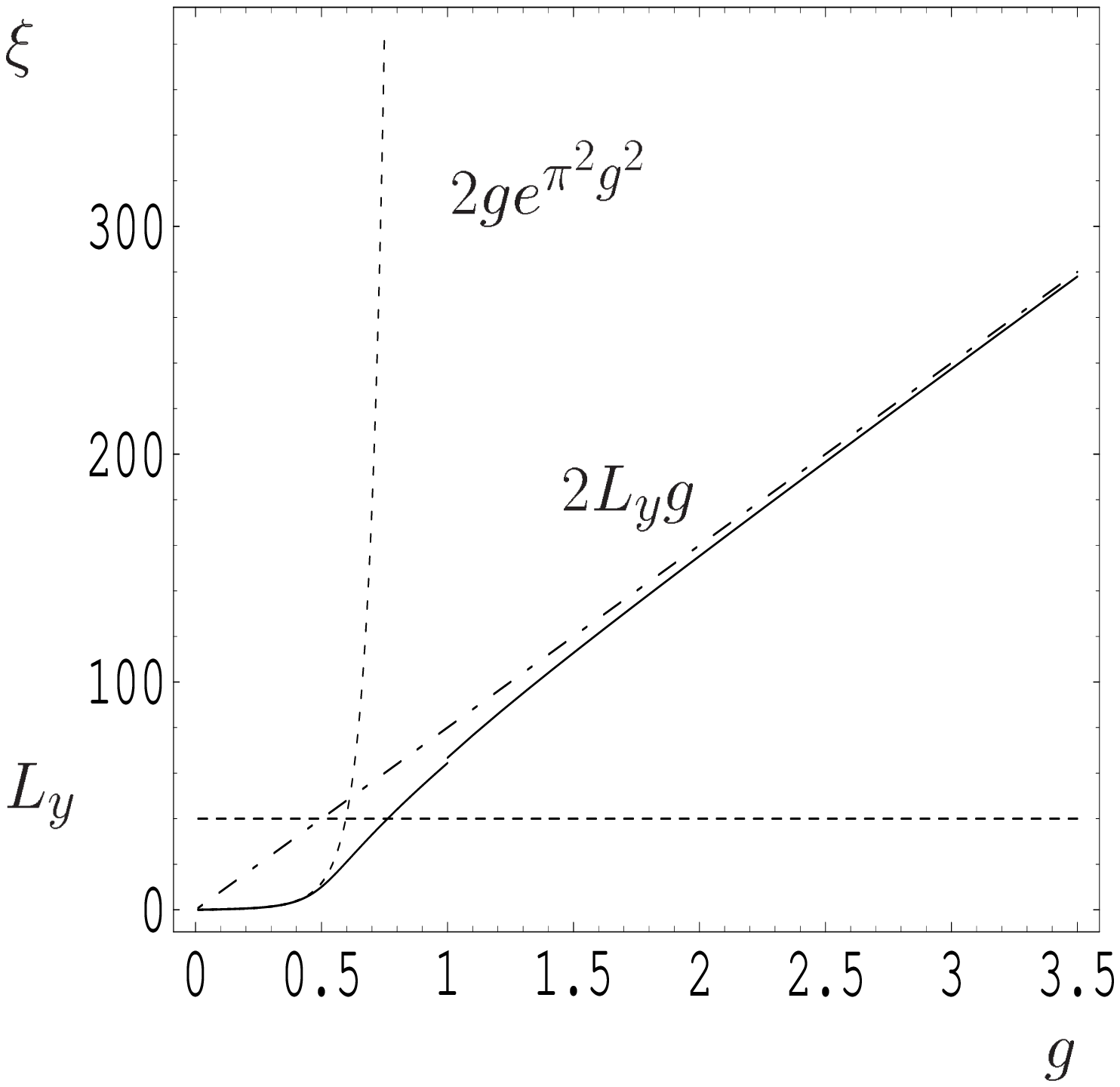}
\vspace{-6cm}
\caption{ The localisation length $\xi$ 
 as function of conductance $g$, for large enough magnetic fields such that  $\xi > L_B$, 
 at fixed width $L_y$.
 For comparison, the quasi--2--dimensional behavior $\xi_{2D} = 2 (g/ k_{\mbox{\tiny F}}) \exp ( \pi^2 g^2 )$
 ( dotted line) 
 and the  quasi--1--dimensional
 limit  $ L_y g$ (dash--dotted line). 
 The width $L_y$ is  indicated by the  dashed line.  
 }
 \label{lcbg}
\end{center}
\end{figure}
\begin{equation} \label{lcblambert}
\xi^2  = L_y^2  \left( \Sigma^2 ( 4 g^2 -1 ) + \frac{2}{ \pi^2} 
   \ln  \left[ \frac{ 1 + (\xi/(2 \pi  L_y) )^2}{1+ (\xi k_0 )^2}
 \right] \right).  
\end{equation}
 Here, the length scale $2 \pi/k_0$ is the   short distance cutoff of the nonperturbative theory, 
 being $l = 2 g(B=0)/k_F$ at moderate magnetic fields, when  $ \omega_c \tau < 1$, 
 and crossing over to the cyclotron length $ l_{\rm cyc}$
 at strong magnetic fields 
 when  $ \omega_c \tau > 1$.

 Here,  $\Sigma =1$, unless  
 the spin degeneracy is broken, and the energy levels are mixed by spin flip 
 as   by scattering from 
  magnetic impurities, then $\Sigma=2$.
 Note that the above results confirm the result for the
localisation length in the  quasi--1--D limit \cite{larkin,dorokhov},
\begin{equation} \label{beta}
\xi \mid_{\xi \gg L_y} = \Sigma \beta g L_y,
\end{equation} 
where $\beta =1,2$ with, without time reversal symmetry. 
 We will set $\Sigma=1$ in the following.    
   We can solve Eq. (\ref{lcblambert}) in two limiting cases. 
 When $L_y \ll \xi$, the quasi--1--dimensional localisation length is obtained with a
 logarithmic correction, 
\begin{equation} \label{lcb1d}
\xi \mid_{\xi \gg L_y} = 2 L_y g \left[ 1 - \frac{1}{\pi^2 g^2} \ln
\sqrt{1+ \left( L_y k_0/(2 \pi ) \right)^2} \right]^{1/2}.
\end{equation}

 In the limit of 2--dimensional localisation, $\xi \ll L_y$,  the
 localisation length  is found to 
 satisfy the equation
\begin{equation}
\xi    \mid_{\xi \ll L_y} = \frac{2 \pi}{k_0} \exp \left[ \frac{\pi^2}{4} ( 4 g^2 -\frac{\xi^2}{L_y^2} )\right]. 
\end{equation}
 Its solution can be written in terms of the
 Lambert--W--function\cite{lambert} as
\begin{equation} \label{lcb2d} 
\xi = \frac{\sqrt{2} L_y}{  \pi }  W_0^{1/2} \left[  \frac{2 \pi^4 }{  k_0^2 L_y^2} \exp (2 \pi^2 g^2 )  \right].
\end{equation} 

 Fixing  the width $L_y$  
 the  dimensional crossover  is  seen  in Fig. \ref{lcbg}, 
 where the localisation length is plotted  as function of the 
 dimensionless conductance g,  using  
   Eq. (\ref{lcb1d})  for $\xi>L_y$, and   Eq. (\ref{lcb2d}) for $\xi < L_y$. 

\section{The magnetic field dependence of the localisation length} 

{\it Weak magnetic field} 

    In  disordered quantum wires  
 without 
 strong spin--orbit scattering  or  magnetic impurities, the electron
   localisation length is enhanced  by 
  a weak magnetic field. 
 If the localisation is quasi--1--dimensional, $\xi > L_y$, then the magnetic field 
 results in a doubling, Eq. (\ref{beta}),  
 $\xi = \beta g L_y $, where $\beta =1, 2$,
 corresponding to no magnetic field, and 
 finite magnetic field, respectively\cite{larkin,dorokhov}.
 Recently, such a doubling of the localisation length was observed experimentally
\cite{khavin}.
  That doubling  is governed by  the magnetic diffusion length 
 $L_B = ( D \tau_B )^{1/2}$, where $\tau_B$ is the 
 magnetic phase shifting time, which is a function of 
 magnetic length $l_B$, mean free path $l$ and width of the wire $L_y$
\cite{mazza}.  
 Thus, the localisation length 
 crosses over to $\xi =2  g L_y $,
 when the magnetic diffusion length 
 becomes smaller than the localisation length, $ L_B < \xi$. 


 This  crossover has recently been derived analytically\cite{prbr,mazza}.
The quasi--1--D localisation length,  
 valid for $\xi >L_y$,  is obtained to be given by: 
\begin{equation} \label{asd1}
\xi ( B) =2  f( L_B/\xi(0)) g L_y. 
\end{equation}
 Here,  
\begin{equation} \label{asd2}
f(x) = 2/(2 +\sqrt{49 + 64 X^2}-\sqrt{25 + 64 X^2}).
\end{equation}
 For a  wire of diffusive width $W>l$, one finds, 
$X =  \xi(0)/L_B =  \pi^2  3^{-1/2} N^2 b$, 
 where $ N_y =  k_{\mbox{\tiny F}} L_y/\pi$ is  the number of transverse channels, 
 and $ b = \omega_c \tau = e B \tau/m$  the dimensionless  magnetic field
 parameter.  The factor $64$ 
 was not noticed in a previous publication \cite{mazza}, and 
is  a consequence of 
 the particular properties of the autocorrelation function 
 of spectral determinants, and its relation to the actual localisation length
 which was used to derive Eqs. (\ref{asd1},\ref{asd2})
 \cite{mazza}. 
  When the localisation is 2--dimensional,  the localisation length $\xi$   becomes exponentially 
 enhanced.
 The crossover between the orthogonal, Eq. (\ref{2d}), and unitary, Eq. (\ref{2db}),  
 localisation length 
is   governed by the parameter $X =  \xi/L_B$. 
Extending the perturbative renormalisation, by calculating the Hikami boxes 
as a function of magnetic field as outlined in appendix B
( This  problem, the magnetic field induced crossover of the weak localisation 
 corrections to the conductivity, has been considered to 2 nd order in 
 $1/g$ also in Ref. \cite{oppermann}, using the complex matrix model),
 one obtains
the following equation for the 2D--localisation length,
\begin{equation} \label{crossover2d}
\xi = l \exp (\pi g ) (1 + X^2)^{1/2-1/(2 \pi g)} \mid_{X = \xi/L_B}.  
\end{equation}

 Thus, when the time reversal symmetry is broken, $\xi > L_B$, the localisation is 
given by  Eq. (\ref{lcb1d})  for $\xi>L_y$, and   Eq. (\ref{lcb2d}) for $\xi <
L_y$, respectively.  
 The effective  short distance cutoff of  the renormalisation is  a function of the magnetic field 
 itself. We get   $2 \pi /k_0= l$ at small magnetic fields, $b <1$, 
 and  $2 \pi/k_0 = l_{cyc}$ at large fields 
  from the renormalisation of  the diffusive nonlinear sigma model. But, we
 should  note that an analysis of the 
 nonlinear sigma model on ballistic scales is needed to get  more reliable information on $1/k_0$.

 {\it Strong magnetic field}

 In the following,   we consider a quantum wire of disordered electrons
 in a strong magnetic field, which is expected to exhibit 
 the quantum Hall effect, 
 when its width $L_y$  and 
the mean free path $l$
do exceed the cyclotron length $l_c$.
 We consider  quantum Hall wires of diffusive width, 
 where the mean free path is smaller than the wire width, 
 $l < L_y$. In the opposite limit, $l> L_y$,
 the ballistic motion between the edges of the wire leads to 
 anomalous  magnetoresistance phenomena due to
 classical   commensurability
 effects  of cyclotron orbits with the confinement potential of the
 wire \cite{roukes,park}.
 This  will not be pursued here,   since
 these effects are  not related to   disorder induced quantum localisation, which is the focus of this article.
 $\sigma_{xx}(B)$ is the conductivity in
  self consistent  Born approximation.
 For weak magnetic field, $b <1$, it is identical to  the Drude result
$\sigma_{xx} = n e^2 \tau/(1+b^2)$. 
 With the electron density $ n_e = E m/(2 \pi \hbar^2 ) $ this
 can be rewritten 
 as 
 $ g(b) = \sigma_{xx}/\sigma_0 = g/(1+b^2)$, 
 where $g =  E \tau/\hbar$ per spin channel. 
Fixing the electron density $n_e$, the Fermi energy depends on magnetic
 field $E_F ( B) $ for $b>1$. In the following,  we will fix the Fermi energy $E_F$, instead. 

 The conductivity  in SCBA for $b>1$, 
 when the cyclotron length $ l_c$ becomes smaller than the 
 mean free path  $l$, or $ \omega_c > 1/\tau$, 
 and disregarding the overlap between Landau bands, 
is given  by  
\begin{equation} \label{scba}
g(B)  = \frac{1}{\pi}  (2 n+1) ( 1 - (E_F-
 E_n)^2/\Gamma^2), 
\end{equation}
for $
 \mid E - E_n \mid < \Gamma$, 
 where $\Gamma^2 = (2/\pi) \hbar^2 \omega_c/\tau$, for
 $\Gamma < \hbar \omega_c$. 

\begin{figure} 
\begin{center}
\includegraphics[width=.48 \textwidth]{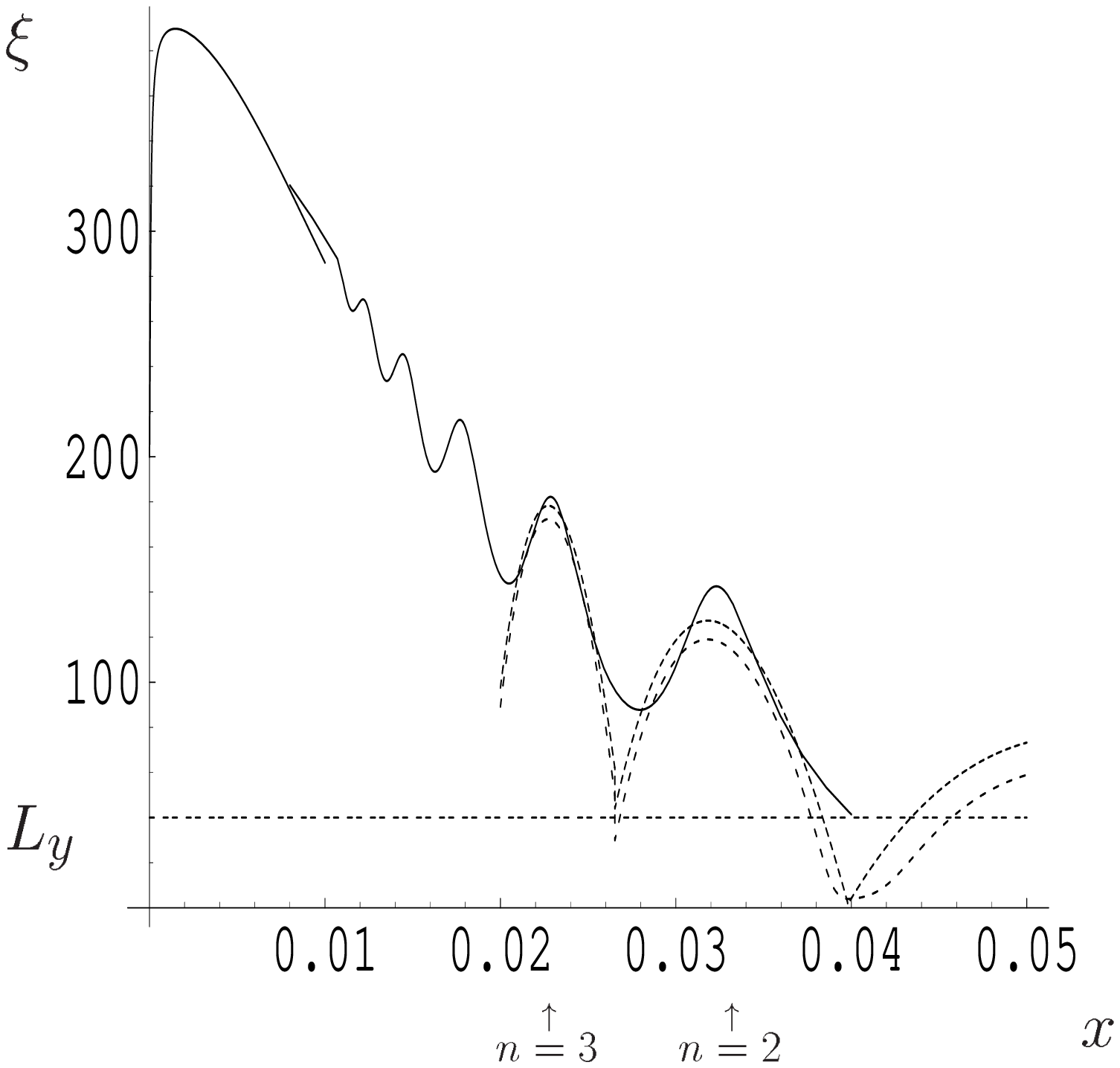}
\vspace{-5cm}
\caption{ 
  The localisation length
 as function of magnetic flux
 through a unit cell of area $a^2= 1/k_F^2$, 
 $x = \frac{a^2}{2 \pi l_B^2}$, 
 with  conductance parameter $g(B=0) = 10$. 
 For weak magnetic field, Eq. (\ref{asd1}) is used. 
 For strong magnetic fields the localisation length is plotted 
  by 
 inserting $g(B)$ in 
 the 2 nd order Born approximation 
 in the formula for the quasi-1-dimensional localisation length, 
Eq. (\ref{1db}),  including a summation over all Landau levels (full line). 
 The short dashed  line is obtained by using the self consistent Born approximation 
 (SCBA) for g(B) for one Landau band, Eq. (\ref{middle}), and inserting it
 into the formula for the quasi-1-dimensional localisation length, 
Eq. (\ref{1db}). 
  The long dashed curve denotes the correpsonding result using 
 the SCBA conductance and inserting it in the 
 crossover  formula  Eq. (\ref{lcb2d}). 
 The width of the wire $L_y = 40 a$ is indicated as the horizontal dashed line. 
 }
 \label{lcbosc}
\end{center}
\end{figure} One obtains thus 
 the localisation length for $b>1$ and $\mid \epsilon/b -n -1/2 \mid <  1$, 
 by substituting the expression for the dimensionless
 conductance Eq. (\ref{scba}), 
 into   Eq. (\ref{lcb1d})  for $\xi>L_y$, and   Eq. (\ref{lcb2d}) for $\xi < L_y$, respectively. 
  Thus, the localisation length is found to oscillate between
 maximal values in the middle of the Landau bands,  and  
 minimal values on the order of the cyclotron  length 
 $l_{\rm cyc}$ in the tail of the Landau bands, as seen 
 in Fig.  \ref{lcbosc}. 

 The localisation is quasi--1--dimensional  as long as $ \xi ( W) > L_y$. 
 We see that for $n>1$ this is,
 for uncorrelated disorder potential, practically always    the case
 in the middle of the Landau  bands,
 with a  logarithmic correction as given by Eq. (10), yielding 
\begin{equation} \label{middle}
\xi_n = \frac{2}{\pi} (2 n + 1)  L_y\left[ 1- \frac{1}{(2 n+1)^2} \ln
 \sqrt{ 1+ \left(\frac{L_y}{l_{\rm cyc}}\right)^2} \right]^{1/2}.   
\end{equation}

 Only in the lowest  Landau band,  Eq. (2) 
 gives a value for the
 localisation  in the middle of the Landau bands 
  $\xi ( B) \approx  (2/\pi)  L_y$, 
 for $n=0$, which  is  smaller than the width $L_y$.
  Thus,   in the lowest  
 Landau band  the  NLSM is 2-dimensional 
 and the topological term becomes fully effective already for 
 small widths $L_y$, see Appendix D.  
 There, the gap in the action of the nonlinear sigma model is 
 diminished due to 
  the presence of the topological term $\sigma_{\rm xy} \neq 0$. 
Then, there are  
 excitations 
  with nonvanishing topological charge $q$, like  Skyrmions
 which  give a phase 
$\exp (i 2 \pi q \sigma_{\rm xy})$ to the functional integral. 
 For  $\sigma_{\rm xy} = (2 n +1)/2$, 
the skyrmions  with $q = \pm 1$ 
give the opposite sign to  the functional integral, 
 $ \exp  (i \pi) = -1$ than configurations without topological charge, 
 $ q =0$, 
 like  vortices\cite{pruiskenrev}. 
  Thus, 
 the presence of the topological term   leads
 to quantum criticality, and a diverging localisation length,
 there. For a wire of finite width, the localisation length then 
 saturates to its critical values 
 which was found numerically to be $\xi_c \approx 1.2 L_y$\cite{hucke}, 
 while analytical estimates exceed this value \cite{meraikh}.

  We note that we have assumed in the derivation 
 of Eqs. (\ref{crossover2d},\ref{middle}),
 that the conductivity is homogenous. 
 In a strong magnetic field, the formation of edge states
 can  result in strongly inhomogenous and
 anisotropic conductivity, thereby  preventing the mixing of 
  edge states with the bulk states\cite{shkledge,schweitzer}. 
 The consequences of these effects will be discussed 
  in more detail, in the next chapter and  
 in a forthcoming article,  including a  numerical analysis\cite{mesoqh}.

 In a real sample there exists a slowly varying potential 
 disorder which is known to stabilize the edge states
 against mixing with bulk states and back scattering \cite{ohtsuki,hajdu}.
   But in the middle of the Landau bands there is always scattering from 
 the  edge states into the bulk \cite{shkledge},  
 so that quasi--1--dimensional localisation\index{localisation} should occur
 for large aspect ratios of the quantum hall bar, $L \gg L_y$. 




\section{ Scaling Function}

 In the above  analysis of a disordered  wire in a magnetic field 
 we disregarded the effect of the topological term, which appears in 
 a derivation of the nonlinear sigma model\cite{pruiskenrev,pruisken,qhcritical},
 see Eq. (\ref{free2db}).  
 It is known that in the two-dimensional limit this 
 topological term is needed in order that the field theory becomes critical 
 in the middle of Landau bands, and the quantum--Hall--transition 
 from localised states in the Landau band tails to a critical 
 state in the middle of the Landau bands can be described\cite{pruiskenrev,pruisken,qhcritical,affleck,tsai,tsvelik}. 
  Can one still learn something about the quantum Hall transition from knowing
 the noncritical  localisation length as function of magnetic field, 
 $\xi(B)$,  as derived above, 
 as function of the wire width $L_y$?  
\begin{figure} 
\begin{center}
\vspace{-1.5cm}
\hspace{-1.2cm}\includegraphics[width=.55 \textwidth]{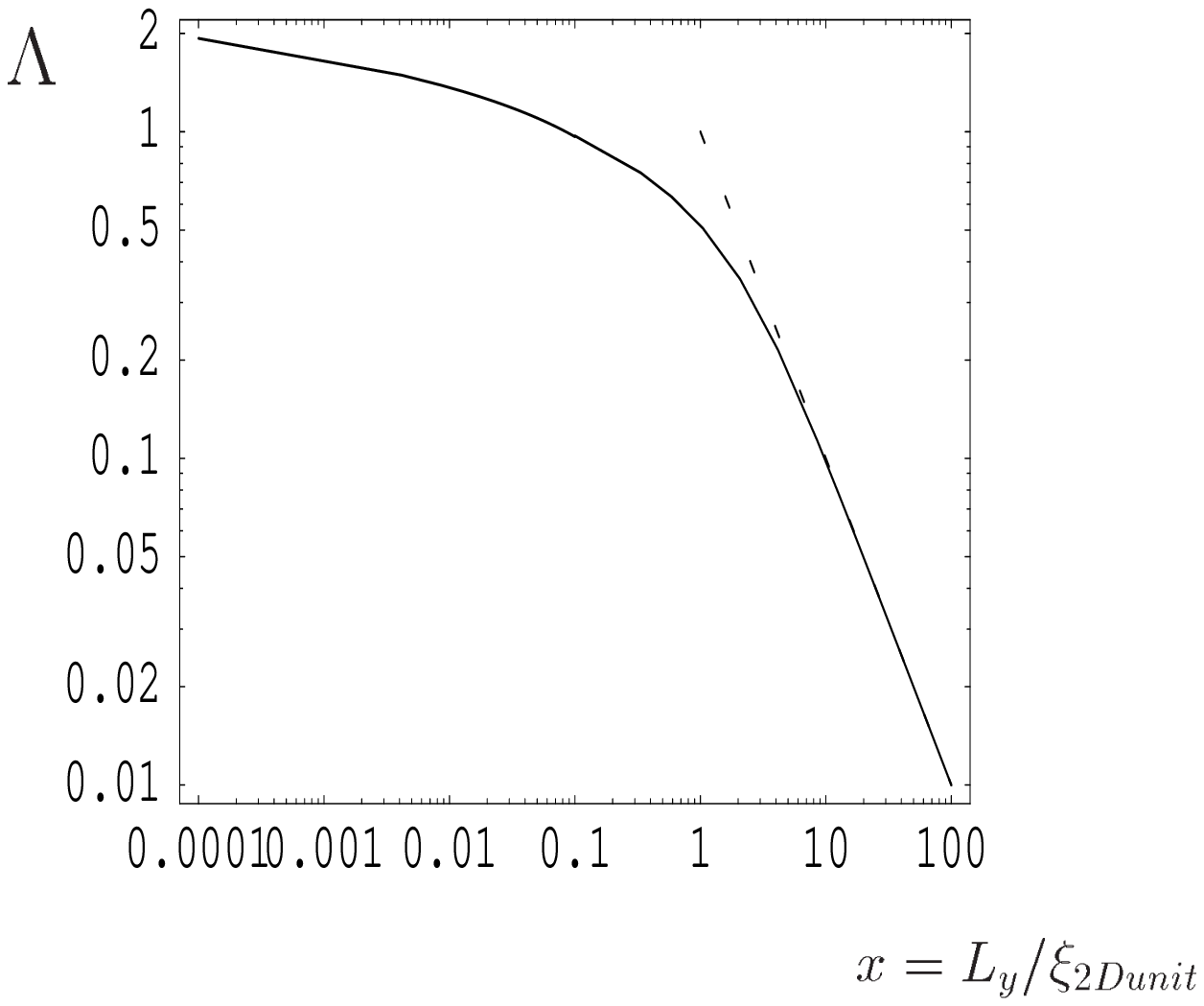}
\vspace{-7.5cm}
\caption{ The  noncritical  scaling function obtained analytically (full line), Eqs. 
(\ref{small},\ref{large}), for the Landau band, $n=3$, with $g_n=7/\pi$.
 The dashed line, $1/x$,  is approached in the 2 D limit, 
 when the localisation length becomes equal to 
     $\xi_{2 D unit}$.  }
\label{scaling}
\end{center}
\end{figure}
 Both in  numerical calculations \cite{hucke} and experiments 
 one needs to perform  a finite size scaling analysis in order 
 to extract the critical divergence of the localisation length, 
 $\xi \sim (E-E_n)^{-\nu}$, when approaching the middle of a Landau band, 
 $E_n$.
  The procedure is to find numerically the scaling function 
$\Lambda = \xi/L_y = \Lambda ( L_y/\tilde{\xi} (E) )$, 
 rescaling with the  critical localisation length, 
 which diverges according to $\tilde{\xi}(E) \sim (E-E_n)^{-\nu}$,
 and does not depend on width $L_y$.  
  Then, one  can determine $\nu$ by optimizing the accuracy of  scaling. 
  The scaling  function is not known a priori. 
 It is clear  that $\Lambda(x) \rightarrow 1/x$ for $x \gg 1$, 
 since in the tails of the Landau band $\xi \ll L_y$,  and
 $\xi$ becomes  independent 
 of $L_y$, approaching $\tilde{\xi}$.

 For  the higher Landau bands, $n>0$,
 the single parameter scaling is not accurate, and 
  it is 
 important  to include also irrelevant  scaling parameters\cite{wegner72}, 
 apart from the relevant parameter, $L_y/\tilde{\xi}(E)$ \cite{hucke,evers}.
 So far the irrelevant scaling parameters have been included in the 
 numerical scaling analysis on a phenomenological ground,  without 
 a precise knowledge of their physical  origin. 
  It has been observed, however, that the irrelavent scaling length increases by orders 
 of magnitude  in higher Landau bands,  for uncorrelated disorder\cite{hucke}.  
 
  Therefore, it seems worthwhile to first analyse,  if the noncritical 
 width dependence  of $\xi(B)$, derived above, Eq. (\ref{lcgeneral}), 
  can yield  analytical knowledge on the scaling function $\Lambda(x)$
 and  moreover  account for the observed irrelevant scaling parameter
 in higher Landau bands. Only then will we try to implement the topological 
 term in the derivation of the scaling function at the end of this section. 
 
 According to Eq. (\ref{large}), the ratio $\Lambda$ 
 scales with the large length scale $\xi_{2 D unit}$, 
 which  is a huge length scale in the middle of higher Landau bands, 
 where $g \gg 1$. Therefore, it is natural to expect that 
 $ \xi_{unit}$ can be identified with the irrelevant length scale $l_{irr}$, 
 and to compare the scaling  function,   Eqs. (\ref{small},\ref{large}), 
 with the one obtained numerically, Eq.  (\ref{irrelevantscaling}). 

\begin{figure} 
\begin{center}
\vspace{-1.5cm}
\hspace{-1.2cm}\includegraphics[width=.55 \textwidth]{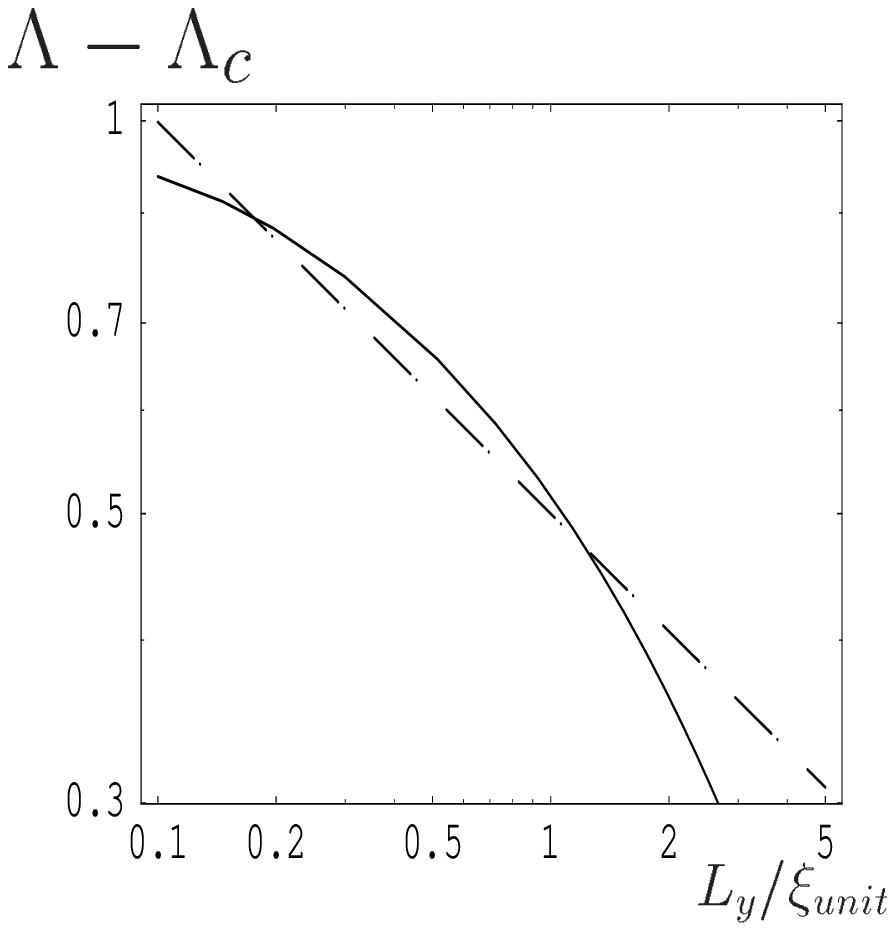}
\vspace{-8.5cm}
\caption{ The irrelevant scaling function obtained analytically
 for the Landau band $n=1$, $g=3/\pi$ (full line) 
 as compared with the function Eq. (\ref{irrelevantscaling}) (dashed line), 
 for $\gamma=.3$ and $c=.5$, 
 double logarithmic scale
.   }
\label{irrfig}
\end{center}
\end{figure} 
 In 
 Fig. \ref{scaling}, we plot the scaling function, as obtained 
  above, Eqs. (\ref{lcb1d},\ref{lcb2d}), as function of 
 $x=L_y/\xi_{2 D unit}$, 
 where $\xi_{2 D unit}$ is the 2D limiting value of the 
 unitary localisation length, 
\begin{equation} \label{xiunit}
\xi_{2 D unit} = \frac{2 \pi}{k_0} \exp ( \pi^2 g^2 ). 
\end{equation}
 Thus, Eqs.  (\ref{lcb1d},\ref{lcb2d}) become 
\begin{equation} \label{small}
\Lambda = \frac{2}{\pi} \sqrt{- \ln \sqrt{ x^2 + ( 2\pi/ k_0\xi_{2 D unit})^2 } }, 
\end{equation}
 for $x= L_y/ \xi_{2 D unit}  <  \exp (-\pi^2/4) \approx .085 $, 
 while  
\begin{equation} \label{large}
\Lambda = \frac{\sqrt{2}}{\pi} \sqrt{W_0 \left( \frac{\pi^2}{2}
 \frac{1}{x^2} \right)}, 
\end{equation}
 for  $x= L_y/\xi_{2 D unit} > \exp (-\pi^2/4)$. 
 This noncritical scaling function  
 is plotted in Fig. \ref{scaling}, where $g=7/\pi$ has been chosen, 
 corresponding to the unrenormalised conductance in the Landau band, $n=3$.
 We note that the derivation is only valid for $g \gg1$, so that 
  we are  able to compare  this function only with the scaling function 
 in higher Landau bands, where $g_n \gg 1$.
 It is expected that this noncritical scaling function 
is accurate as long as   $x= L_y/\xi_{2 D unit} < 1$. 
 
 Close to  the critical point,  $x= L_y/\xi_{2 D unit} \gg 1$, 
  the  scaling function is from the numerical analysis 
 obtained to be in the middle of the Landau band   
\begin{equation} \label{irrelevantscaling} 
\Lambda - \Lambda_c = \frac{\xi}{L_y} = c (\frac{L_y}{l_{irr}})^{-\gamma}, 
\end{equation}
 where $\Lambda_c = 1.2$, 
 and the irrelevant critical
 scaling exponent 
is numerically found to be 
$\gamma = .38 \pm .04$, and  $c$ is a constant. 
 
 In  Fig. \ref{irrfig}, we plot 
  $\Lambda - \Lambda_c$, Eq. (\ref{irrelevantscaling}) 
 using $\gamma = .3$ and  $c = .5$,  
 and  compare it  with $\Lambda$ 
 as obtained from the result of the noncritical analysis, 
 Eq. (\ref{large}). Note that this function converges to zero as $L_y
 \rightarrow \infty$, 
 corresponding to  $\Lambda_{\rm crit}=0$, 
since it was obtained from the noncritical field theory, 
 disregarding the topological term. Thus, as expected,  the form 
 of the noncritical  irrelevant scaling function defers from  
 the results of the numerical analysis as seen in Fig. (\ref{irrfig}), 
 but is in some quantitative agrrement.  
  Inspite of this,  it is expected that the scaling function itself 
 is changed by the presence of the critical point in the middle 
 of the Landau band, and the  similarity to 
 the noncritical scaling function derived above,  
 is only of  qualitative nature. 
 Since in higher Landau levels one is in the study 
 of wires of finite width $L_y$ for uncorrelated disorder 
 always far away from  the critical 
 point, as shown above, we conclude that 
 this  noncritical scaling  function is important in order to 
 enable one  to analyse
 the quantum Hall transition in higher Landau bands, 
 $n>1$. 
 We can estimate the region of criticality by the 
 condition that $\tilde{\xi}(E)/\xi_{2 D unit} > 1$. 
 Thereby we find for the interval of criticality around 
 $E_n$: 
\begin{equation}
\Delta E = \Gamma \exp ( - \frac{(2 n + 1)^2}{\nu} ),
\end{equation} 
 which yields $\Delta E/\Gamma = .65, .02, 2. ~ 10^{-5},...$ for 
 $n=0,1,2,...$. 
 Thus, we conclude that criticality can only be observed in the lowest two Landau levels. 
 Since the width of the quantum Hall plateaus is determined by the condition 
$\xi(E) = L_y$,
 there are nevertheless wide plateaus between higher Landau bands, and 
 we conclude that criticality is not essential to observe the 
 quantum Hall effect. 

  Next, let us consider the effect of 
 the topological term in the derivation of the scaling function.  
 At small length scales, in high Landau bands, 
 the dimensionless conductance $\sigma_{xx}$ is large, and the instanton approximation 
 can be used. 
 To this end, one finds solutions which minimize the action of the NLSM, 
Eq. (\ref{free2db}),
\begin{eqnarray} \label{free2dbtop}
F &=& \frac{h }{16 e^2}  
\int d {\bf x} \sum_{i=x,y}  \sigma (\omega=0)_{ii} ({\bf x})
 STr \left[ ( { \nabla_i} Q({\bf x} ))^2 \right] 
\nonumber \\ 
&-& \frac{1}{16} \frac{h}{e^2}   \int d {\bf x}  \sigma (\omega=0)_{xy}
({\bf x}) STr \left[ Q \partial_x Q \partial_y Q \right],
\end{eqnarray}
where $ \sigma (\omega=0)_{xy} =
 \sigma^I (\omega=0)_{xy}
({\bf x}) + \sigma^{II} (\omega=0)_{xy}$ 
 where $ \sigma^I (\omega=0)_{xy}$ is the dissipative part of the 
 Hall conductivity\cite{ando,pruisken,hajdu}, and
$\sigma^{II} (\omega=0)_{xy} = - e dn/dB $, 
where $n$ is the particle density, 
which yields  a finite 
 contribution at the boundary of the 
 wire in the presence of a confinement 
 potential, from  the edge states \cite{streda,hajdu,ef}.

 Disregarding the spatial variation of the coupling functions
  $\sigma_{ij} ({\bf x})$ in Eq. (\ref{free2dbtop}), and 
assuming isotropy, 
  one  founds that there are instantons with 
 nonzero topological charge $q$, which 
 are identical to the skyrmions of the compact $O(3)$ NLSM, 
 as obained form  the compact part of the superymmetric NLSM\cite{pruisken,ef}. 
 Their action is given by 
 \begin{equation}
F_{q} = 2 \pi \mid q \mid \sigma_{xx} + 2 \pi i q \sigma_{x y}, 
\end{equation}
where $\sigma_{xx} = \sigma_{yy}$ and $\sigma_{x y}$ are the spatially averaged 
 conductivities. 
 Now, we can repeat the derivation of the scaling function, 
 by integrating out Gaussian fluctuations around these instantons. 
 It is clear, however that the contribution from instantons with 
 $q \neq 0$ is negligible, as long as $\sigma_{xx} > 1$. 
  Within the validity of the $1/g$ expansion one does not find
a sizable  influence of the topological term  on the scaling function
 $\Lambda = \xi/\L_y$.  
 Still,  the tendency is seen that at $\sigma_{xy}=1/2$ the, 
 renormalisation of the longitudinal conductance is slowed down
 and one may conclude  from this observation the two parameter 
scaling diagram with a critical state of finite conductance
 $0 <\sigma^* < 1$ \cite{khmelnitskii,pruisken}.

   Taking into account  the spatial variation of $\sigma_{ij} ({\bf x})$, 
 there are extended regions where  $\sigma_{yy} ({\bf x})\rightarrow 0$,
 indicating the decoupling 
of  the edge states from the bulk states\cite{shkledge}.  
  Thus, the free energy for spatial variations of 
 $Q$ is reduced in these regions. Thereby, one can find instantons 
 with nonzero topological charge $q$, whose spatial variations
 are restricted to these edge regions with vanishing real part of the 
 free energy: $F_{q ~ {\rm edge}} = i 2 \pi \sigma_{xx} ( {\rm edge} )$, 
 where $\sigma_{xx} ( {\rm edge} )$ is the Hall conductance of the edge 
 states, which is quantised to   integer values. 
  Thus, we conclude that the renormalisation and thereby the scaling function 
 of the bulk, $\Lambda = \xi/L_y$, is not  influenced
noticeably  by the presence of the edge states for $g>1$.
 
 Closer to the critical point, the NLSM, Eq. (\ref{free2dbtop}),
can not be used to derive further information, since that theory
 flows to  strong coupling, $g < 1$.  
It has been established numerically that the quantun Hall criticality
 is not sensitive to the type of disorder. This observation found 
 further support by the proof that 
 the Hamiltonian of a  chain of antiferromagnetically 
 interacting superspins can be derived both  from 
 the nonlinear sigma model
  for short ranged disorder at $\sigma_{xy} =1/2$\cite{zirnchain},  
as well as from the Chalker--Coddington model
 \cite{lee}, 
 which is the reduced version of the quantum percolating network 
 model of unidirectional (chiral) drifting modes along equipotential lines
 of a slowly varying disorder potential\cite{coddington}. 
It has been shown by numerical solution of    
 a finite number of antiferromagnetically coupled super spins that this theory is
 critical.  So far, no analytical  information has been 
 obtained for the critical parameters, such as the localisation 
 exponent, $\nu$ and the critical value $\Lambda_c$. 
  However, building on this model of a superspin chain, 
   supersymmetric conformal field theories
 have been suggested, which ultimately should yield the 
 critical parameters of the qunatum Hall transition\cite{conform1,conform2}. 
 The  critical value of the scaling function, $\Lambda_c$ has been 
 related to   the 
 free parameters of the conformal field theory \cite{conform2}. 
  Restricting this  theory to quasi-1D, by choosing a finite width 
$L_y$, on the order of $\xi_{2D unit}$, which serves as the 
 ultraviolet cutoff of the conformal field theory, 
 one finds that the critical value of the scaling function 
 $\Lambda_c$ is inversely proportional to the 
 gap between the lowest two   eigenvalues of the Laplace-Beltrami operator
 of this reduced class of  supersymmetric conformal field theories. 
  Thus, it seems that the critical value $\Lambda_c$
 of the  scaling function at  the critical point can be obtained from 
 the dimensional crossover of the conformal field theory.   
So far, however, the critical exponent of the localisation length has not been 
 derived from the conformal field theory, nor from the theory of superspin chains.
 Therefore, it is  so far not possible to give an anlytical derivation of the 
 scaling function close to criticality. 

\section{ Conclusions }

  In disordered quantum wires the electrons are localised
 due to quantum interference 
 along the wire 
   with 
  a localisation length which scales linearly with the wire width, 
 as long as  the electrons can diffuse freely across the wire width.  
   For wires, which would classically be good  metals, as 
  characterised  by a large dimensionless conductance
 $ g = k_F l \gg 1$  the 2D 
 quantum localisation limit   is never reached, but rather a slow crossover
 between quasi--1--D and 2--D localisation  occurs as function of the 
 wire width.  Therefore, we think that the crossover function derived 
 here, can be relevant for the study of strong localisation
 in weak magnetic fields in disordered quantum wires. 
 These  have been studied mainly by means of activated transport
 measurements \cite{khavin}. 
  Capacitance measurements would yield the localisation length 
 directly. Since insulators are dielectrics, with dipole moments 
 proportional to their localisation length, 
 the metallic divergence of the 
 dielectrical constant is cutoff, 
  $ \epsilon (q \rightarrow 0) \sim  \xi^2$.
In general, for an insulator one obtains  
 for $T \ll \Delta_c = 1/\xi^d \nu_d$, 
\cite{reviews1}
\begin{equation} 
 \epsilon (q \rightarrow 0, \omega = 0) = 4 \pi e^2 \frac{dn}{d \mu} \xi^2.
\end{equation}
 Thus, the measurement of the dielectrical constant has been   
 used  to study the metal--insulator transition\index{metal--insulator transition}\cite{reviews2}, where the localisation length
 and thus the dielectrical constant is diverging\cite{milligan}. 
For a quasi--1--dimensional wire one obtains
\begin{equation}
  \epsilon (q \rightarrow 0, \omega = 0) = 32 \zeta (3) e^2 \nu_d   \xi^2,
\end{equation}
 where $\zeta$ is the Riemann $\zeta$--function  \cite{larkin}. 
  Measuring the  
 magnetocapacitance, $C(B) = \epsilon_0  \epsilon(B) S/L$, 
 where $S$ is the cross section and $L$ the length of the wire, 
   one  would expect an enhancement of  
 the
 dielectrical constant  $ \epsilon(B) $\index{$ \epsilon(B) $} by a factor 4 
 as the magnetic field is turned on. 
 To our knowledge this positive magnetocapacitance in a wire has not yet been 
 experimentally observed, and would be a means to study the 
 dimensional crossover of localisation directly. 

 In a strong magnetic field, the kinetic energy is quenched, 
 resulting in enhanced localisation. While in the
 tails of the Landau bands the localisation length is small, 
 on the order of the cyclotron length, 
 it increases towards the center of the Landau bands, due to an increased
 classical conductance. For wires of finite width, this results in 
 a dimensional crossover of localisation form 2-- to 1-- dimensional behavior. 
 The noncritical crossover function derived above is relevant for 
 localisation in higher Landau bands, where the noncritical 
 2D--localisation length  is exponentially large, dominating its behavior 
 since  the critical point in the middle of the Landau band 
 becomes relevant only in the 2D limit. 
 
 In the tails of the Landau bands, extended edge states 
 exist due to the edge confinement potential of the wires, 
 which can carry a quantised Hall current. When the dimensinal 
 crossover of the localisation of the bulk states occurs, 
 the edge states are expected to mix and become localised along the wire. 
 In order to study this localisation transition, the edge states have to be
 taken into account explicitly, by accounting for a strongly inhomgenous and 
 anisotropic conductivity. The  ballsitic length scales of the edge states
  exceeding the 
 elastic mean free path in the bulk, do have to be taken into account 
 explicitly, 
 in the derivation of the field theory of localisation, 
 as outlined in appendix D. 
  A full analysis of this theory, including the 
 edge states and the topological term, in deriving 
 dimensional 
 crossover of localisation remains to be done, as well as a numerical 
 analysis of the metal-insulator transition of the edge states 
in quantum Hall wires\cite{mesoqh}.

\label{sec:summary}

\noindent

\section*{ACKNOWLEDGMENTS}
The author gratefully acknowledges
 usefull discussions with
Bodo Huckestein, Tomi Ohtsuki,
 Bernhard Kramer,   Mikhail Raikh, Ferdinand Evers, 
Alexander Struck and  Isa Zarekeshev.
 This research was supported 
by the German Research Council (DFG),  Grant No. Kr 627/10,
 and under  the 
   Schwerpunkt  "Quanten--Hall--Effekt", 
 as well as  by  EU TMR--networks
 under Grants. No. FMRX--CT98--0180 and HPRN--CT2000--0144.

 \appendix 

\section{}
 Information on the 
  dimensional crossover in a wire of finite 
 width $L_y$,  can be obtained 
 from  the  renormalization of the
 action of the nonperturbative theory of disordered electrons,
 the nonlinear sigma model,  Eq. (\ref{free2d})\cite{weg}.

  First, let us consider the problem 
 without magnetic field, $B=0$.
 The coupling parameter is  the conductance per spin channel, 
$g =\sigma_{xx}/\sigma_0$ in 
 the action  for $B=0$, which  is  given by 
 \begin{equation} \label{free2d}
F = \frac{g }{16}  
\int d {\bf x}  
 STr[ ( { \bf \nabla} Q({\bf x} ))^2].  
\end{equation}
  Going to momentum representation, one performs 
 successive integration, over modes with
 momenta within the interval $k_0/b^l 
< \mid {\bf k} \mid  < k_0/b^{l-1}$, 
 where $k_0 \sim 1/l$ is the high momentum cutoff
of the diffusive NLSM Eq. (\ref{free2d}). 
  $b >1$, is    the renormalisation parameter. Rescaling the 
 coupling parameter $g$ after each renormalisation step $l$, 
 integer, one obtains in one loop approximation,  
\begin{equation} \label{rg}
g  \rightarrow
 \tilde{g} =
 g  ( 1 - \frac{2}{g}
\int_{0 < \mid {\bf k} \mid < k_0 }
 \frac{d {\bf k}}{(2 \pi)^2}
\frac{1}{{\bf k}^2 + \lambda^2}),
\end{equation}
 where $\lambda$ is the low momentum cutoff. 
 The  first  order term in the  perturbative renormalization 
 in $1/g$ corresponds to
  the weak localisation correction to the conductivity.
 One can estimate  the localisation length
 $\xi$, by the fact that  the  conductivity
 of a wire of length $\xi$ is unity, 
   $\tilde{g}  \rightarrow  1$,
  when  $\lambda =1/\xi$.
 Noting that 
$$\int_{0 < \mid {\bf k} \mid < k_0} 
 \frac{d {\bf k}}{(2 \pi)^2}
 \rightarrow \frac{1}{L_y}\sum_{n_y} \int \frac{ d k_x}{2 \pi},$$
 for a wire of finite width $L_y$, with $ k_y = 2 \pi  n_y/ L_y$, 
 where $n_y$ is an integer, one finds that the 
 localisation length in a wire of finite width $L_y$ 
 satisfies   the equation
\begin{eqnarray}
\xi &=&  g W
\nonumber \\ 
 &-& \frac{2}{\pi^2} L_y \sum_{n=1}^{N_0} \frac{1}{\sqrt{n^2 + ( N_0/(k_0 \xi))^2}}
\nonumber \\ 
&& \arctan \left( 
 \frac{N_0}{\sqrt{n^2 + ( N_0/(k_0 \xi))^2}}\right) ,
\end{eqnarray}
 where $ N_0 = k_0 L_y/(2 \pi) $.
  For $N_0 \gg 1$ this equation can be approximated by  Eq. (\ref{lcgeneral}). 



\section{}

\begin{figure} 
\begin{center}
\includegraphics[width=.44 \textwidth]{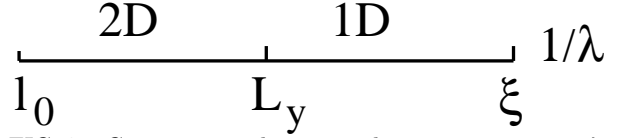}
\caption{ Crossover in dimensionality as momentum $\lambda$ of renormalisation is
  changed.  }
\label{crossoverscale}
\end{center}
\end{figure}

 In a finite magnetic field, the first order
 in $1/g$  correction to the conductance is vanishing. 
 An efficient way to do the perturbative renormalisation to second order
 in $1/g$  is, to start from the supersymmetric 
 nonlinear sigma model and do an expansion around its 
 classical point, 
 as done in   Ref. \cite{ef}
 for the pure  unitary limit. 
 Here we extend this derivation taking into account the finite 
 wire width $L_y$. 
  We note that the dimensionality changes as one integrates out  
 the  $Q-modes$ from large momenta, corresponding to
 the smallest length scales, which is $l_{\rm 0}$ in the  unitary limit, 
 to the largest length scale, which is the localisation length $\xi$, see
 Fig. (\ref{crossoverscale}). 

 Integrating the renormalisation flow from the smallest to the largest 
 length scale, one finds for $\xi > L_y$: 
\begin{eqnarray} \label{ren} 
 4 \int_{g(l_0)}^{g(\xi)} d g g &=& \left[ 16 \lim _{\delta \rightarrow 0} 
 \left(\frac{1}{2} - \frac{1}{2-\delta} \right) \left( I_{2D L_y}^2- l_{2D l}^2 
\right) \right.
\nonumber \\
   &+& \left.  16 \left( \frac{1}{2} -1  \right)
 \left( I_{1D \xi}^2 - I_{1D  l}^2 \right)  \right].  
\end{eqnarray}

 Here, 
\begin{equation}
I_{2D x} =  \int_{1/x < k < \infty} \frac{d {\bf k}   }{ (2 \pi)^2 }
 \frac{1}{k^2  +  \lambda^2},
\end{equation} 
and 
\begin{eqnarray}
I_{1D x} &=&  \int_{1/x < k_x < \infty} \frac{d  k_x}{  \pi L_y }
 \frac{1}{k_x^2  +  \lambda^2}
\nonumber \\
&=& \frac{1}{\pi L_y \lambda}
\left( \frac{\pi}{2} - \arctan \frac{1}{x \lambda}
 \right).  
\end{eqnarray}

 Clearly, we cannot simply set $ \delta=0$ in the first  term
 of Eq. (\ref{ren}), 
 because of the logarithmic divergency of the integral $I_{L_y,l_0}$.  
 Going to dimension $ d = 2 - \delta$, and taking the limit  $ \delta
 \rightarrow 0$, 
   the expression, one has to  evaluate, is given by, 
\begin{equation}
K = \lim_{\delta \rightarrow 0} \left( \frac{1}{2} 
- \frac{1}{2 - \delta} \right) \left( \int \frac{d \Omega_{2 -\delta}}{(2
\pi)^{2-\delta}} \right)^2
 \left( \int d k \frac{k^{1-\delta}}{k^2 + \lambda^2} \right)^2, 
\end{equation}
 where $ \int  d \Omega_{2 -\delta}$ is the angular integral 
 in $ 2 - \delta$ dimensions.  
 By performing an   analytical continuation, 
\begin{equation}
 \int_0^{\infty} 
 d k \frac{k^{1-\delta}}{k^2 + \lambda^2} = \frac{\pi i}{(-\lambda^2)^{\delta}}
\frac{1}{1- \exp ( -  2 \pi i \delta )},
   \end{equation}
  and  using that
 $ \lim_{\delta \rightarrow 0} \left(- \frac{1}{\delta } k^{- \delta}
\right) = \ln k$, one finds: 
\begin{equation}
 K = \frac{1}{8 \pi^2} \ln \lambda.  
\end{equation}
 Thus, setting $\lambda = 1/\xi$, 
 we get, that the localisation length satisfies Eq. (\ref{lcblambert}),  
\begin{equation} \label{lcblambert1}
\xi^2  = L_y^2  \left( 4 ( g^2 -1 ) + \frac{2}{ \pi^2} 
   \ln  \left[ \frac{ 1 + (   2 \pi \xi/L_y )^2}{  (   2 \pi \xi/L_y )^2 + (\xi k_0 )^2}
 \right] \right). 
\end{equation}

 
\section{}
 Here, we   extend the  derivation
 of the localisation length in the 2 D limit
 to the crossover in a 
 magnetic field between the orthogonal and unitary limits. 

 One obtains in two loop approxmation, 
\begin{eqnarray} \label{rg}
 \tilde{g} &=&
 g \left\{ 1 - \frac{4}{g}
\int_{0 <  { k}  < k_0 }
 \frac{d {\bf k}}{(2 \pi)^2}
\frac{1}{{ k}^2 + \lambda^2 + 1/ D \tau_B } \right.
\nonumber \\
&+&  \frac{16}{g^2} ( \frac{1}{2} - \frac{1}{d} ) 
 \left[ ( \int_{0 < \mid {\bf k} \mid < k_0 }
 \frac{d {\bf k}}{(2 \pi)^2}
\frac{1}{{ k}^2 + \lambda^2  })^2 \right. 
\nonumber \\
 &-& \left. \left. ( \int_{0 <  { k}  < k_0 }
 \frac{d {\bf k}}{(2 \pi)^2}
\frac{1}{{ k}^2 + \lambda^2 + 1/ D \tau_B })^2
 \right]
\right\},
\end{eqnarray}
where $\lambda$ is the low momentum cutoff. 
 Setting
 the lower momentum cutoff equal to the inverse localisation length, 
 $ \lambda = 1/\xi$,  we find   
\begin{eqnarray}
\tilde{g} &=& 
 g \left\{ 1- \frac{2}{g \pi } 
 \left(\ln \left(\frac{\xi}{l}\right) - \ln \sqrt{1 + 
  \frac{\xi^2}{ D \tau_B} } \right) \right.
\nonumber \\
 &-& \left. \frac{1}{  \pi^2 g^2} 
   \ln \sqrt{ 1+  \frac{\xi^2}{ D \tau_B} } 
\right\} \rightarrow  1.  
\end{eqnarray} 
 Thereby one obtains the equation  for the localisation 
 length in a magnetic field, Eq. (\ref{crossover2d}).


\section{}

In the following we review the nonperturbative 
theory of a disordered quantum wire in a magnetic field. 
   The Hamiltonian of disordered noninteracting
 electrons is 
\begin{equation}
H= \left[{\bf p
}- q {\bf A} \right]^2/2m + V({\bf x})  + V_0({\bf x}),
\end{equation}
 where $q$ is the electron charge.  
 $V({\bf x})$ is taken to be a Gaussian distributed random function, with a
 distribution function      
\begin{equation} 
 P( V ) = \exp ( -  \int \frac{d {\bf x}}{Vol.} \frac{d {\bf x'}}{Vol.}
 J ( {\bf x} - {\bf x'}) 
 V ( {\bf x} ) V( {\bf x'}) ).
\end{equation}  
 Impurity averaging is thus given by  $<...>_V =
  \int \prod_{{\bf x}} d V P(V) ... $.
 We take    $$J({\bf x} - {\bf x'}) = Vol. \Delta \hbar/\tau \delta
 ({\bf x} - {\bf x'}) $$ for uncorrelated impurities, 
 where $1/\tau$ is the elastic scattering rate and 
 $\Delta=1/(\nu Vol.)$ the mean level spacing of the mesoscopic sample
 with volume Vol..  
$V_0({\bf x})$ is the electrostatic confinement potential defining the 
 width of the wire $L_y$. 
 The vector potential is used in 
 the gauge ${\bf A} = (- B y, 0, 0 )$, where $x$ is the coordinate
 along  the wire of length L,
  $y$ the one  in the direction perpendicular both 
 to the wire
  and the magnetic field ${\bf B}$,
 which is directed  perpendicular to the wire. 
 The electron spin  degree of freedom is not considered here.

 While the disorder averaged electron wave function
 amplitude decays on time scales  on the order of
 the elastic scattering time $\tau$, 
 information on  quantum localisation
 is contained in 
 the impurity averaged evolution of the electron
  density $n({\bf x},t) = <\mid \psi ( {\bf x}, t) \mid^2>$. 
 Thus, nonperturbative averaging of 
 products of retarded and advanced propagators, $ <G^R ( E) G^A (E')>$ 
 has to be performed to obtain information on quantum localisation.

 In usefull anology to the study of spin systems, 
 the supersymmetry method contracts the information
 on localisation  into a theory of Goldstone modes $Q$,
 arising from the global symmetry  of rotations between 
 the retarded propagator ("spin up")
 and the advanced propagator ("spin down")
 in a representation of superfields
 (composed of scalar and Grassmann components). 
 Spatial fluctuations of these modes contribute to  
 the partition function, 
\begin{equation}
Z  = \int \prod d Q_{4 \times 4}
 ({\bf x}) \exp( - F [ Q ] ), 
\end{equation} 
and are governed by  the action
\begin{eqnarray} \label{exactfree}
F[Q] = && \frac{\pi}{4}
 \frac{\hbar}{\Delta \tau}
  \int \frac{d {\bf x}}{S L} Tr Q_{4 \times 4}({\bf x})^2 )
\nonumber \\ &&
 + \frac{1}{2} \int  d {\bf x}
 <{\bf x} \mid Tr \ln  (
 G ( \hat{x}, \hat{p} ) \mid {\bf x} >,
\end{eqnarray} 
 where
\begin{equation}
 G( \hat{x}, \hat{p} ) = 1/ ( 
\frac{1}{2}  \omega \Lambda_3 - 
\frac{(\hat{p}-q A )^2}{2 m} - V_0(\hat{x}) + i \frac{\hbar}{2 \tau} 
Q_{4 \times 4}(\hat{x}) ).
\end{equation}
To summarize the notation, here, and in the following, 
$\Lambda_i$ are the  Pauli matrices in the subbasis
 of the retarded and advanced propagators. 
 We used the notation $\hat{x}$, in order to stress that it
 is an operator, and  does not commute
 with the kinetic energy term
 $H_0 =(\hat{p}-q A )^2/2 m$.
  Here, $\omega = E-E'$ breaks the symmetry
 between  the retarded and advanced sector.
 The   long wavelength modes
 of $Q$, do contain the nonperturbative   information on the 
 diffuson and Cooperon modes, and thus on localisation. 

 In order to consider the action of these long wavelength 
 modes governing the physics of diffusion and localisation, 
 one  can now expand around the saddle point solution of the action
 of $Q$, $\delta F = 0$, satisfying  for $\omega =0$, 
\begin{equation} \label{saddle}
 Q =  i/(\pi \nu) < {\bf x} \mid 
1/( E - H_0 - V_0({\bf x}) + i/(2 \tau) Q )
\mid {\bf x} >. 
\end{equation}  
 This saddle point equation is found to be 
 solved by $ Q_0 = \Lambda_3 P $, which is the self consistent 
 Born approximation for the self energy P. 
 At $\omega =0$ the rotations $ U$, 
 which leave $Q$ in the supersymmetric 
 space,  yield the complete manifold 
 of saddle point solutions as
 $ Q = \bar{U} \Lambda_3 P U$, where $ U \bar{U} = 1$, 
 with  $Q^T C = C Q$.  In general, in order to account for the ballistic motion 
 of electrons along the edges, or to account for different sources
 of randomness, a directional dependence of the matrix $ U = U ({\bf x}. 
{\bf n}) $ where $  {\bf n} = {\bf p }/\mid {\bf p} \mid$ 
 has to be considered  \cite{taras1,blanter}.
 The modes which leave $\Lambda_3$ invariant are surplus, and 
 can  be factorized out, leaving the 
 saddle point solutions to be elements of the 
 semisimple supersymmetric space 
$ Gl(2 \mid 2)/( Gl(1 \mid 1) \times Gl (1 \mid 1)) $ \cite{zirnbauer3}. 
 In addition to these gapless  modes there 
 are massive longitudinal modes with $Q^2 \neq 1$, which  
 can be integrated out\cite{ef}, and  
 the partition function 
 thereby  reduces to  a functional integral over the  transverse
 modes $U$.

 Now, the action of  finite frequency $\omega$ and spatial fluctuations
 of $Q$ around the saddle point solution can be found 
 by an expansion of the action $F$, Eq. (\ref{exactfree}). 
 Inserting $ Q = \bar{U} \Lambda_3 P U$, 
    into Eq. (\ref{exactfree}), 
 and performing the cyclic permutation 
 of $U$ under the trace $Tr$, yields\cite{pruisken}, 
\begin{equation}
 F = - \frac{1}{2} \int  d {\bf x}
 <{\bf x} \mid Tr \ln  (
 G_0^{-1} - U [ H_0, \bar{U} ] + \omega  U \Lambda  \bar{U} ) \mid {\bf x} >,
\end{equation} 
 where 
\begin{equation}
G_0^{-1} = E - H_0 - V_0({\bf x})  + \frac{i \hbar }{2 \tau} \Lambda P. 
\end{equation}

Expansion to first order in the energy difference 
$\omega$                                      
and to second order in the  commutator 
$U [ H_0, \bar{U} ]$, 
 yields, 
\begin{eqnarray}\label{free2}
F[U]  &=&
- \frac{1}{2}  \omega \int  d {\bf x} <{\bf x} \mid
 Tr G_{0 E} U \Lambda \bar{U} \mid {\bf x} >
 \nonumber \\ &+&
\frac{1}{2}\int  d {\bf x} <{\bf x} \mid Tr G_{0 E} U [ H_0, \bar{U} ] \mid {\bf x} >
\nonumber \\ &+&
\frac{1}{4}\int  d {\bf x} <{\bf x} \mid Tr (G_{0 E} U [ H_0, \bar{U} ])^2 \mid {\bf x} >.
\end{eqnarray}

The first order term in 
$U [ H_0, \bar{U} ]$ is proportional to the local current, 
 and found to be finite only at the edge of the wire in a strong magnetic 
 field, due to the chiral edge currents.
It can be rewritten as 
\begin{equation}
F_{xy II} = - \frac{1}{8} 
\int dx dy  \frac{\sigma^{II}_{xy}({\bf x})}{e^2/h}   STr Q \partial_x Q \partial_y Q,  
\end{equation} 
 where the prefactor is the nondissipative  term in the Hall conductivity in 
 self consistent Born approximation\cite{hajdu}:
\begin{equation}
\sigma^{II}_{xy} ({\bf x}) = -\frac{1}{\pi} 
\frac{ \hbar e^2}{m^2} < {\bf x} \mid  ( x \pi_y - y \pi_x ) 
 Im G^R_{E}  \mid  {\bf r} >.
\end{equation}
 One can separate the physics on different length scales, 
 noting that the physics of diffusion and localisation 
 is governed by spatial variations
 of $U$ on length scales larger than the mean free path 
 $l$. The smaller length scale physics, is then included in the 
 correlation function of Green's functions, being related to the 
 conductivity by the  Kubo-Greenwood formula, 
\begin{eqnarray}\label{kubo}
\sigma_{\alpha \beta } ( \omega,  {\bf x} )& =& \frac{\hbar}{\pi S L}
< {\bf r} 
 \mid  \pi_{\alpha}  G^R_{0 E} {\bf  \pi_{\beta} }   G^A_{0 E+\omega} \mid {\bf r} > , 
\end{eqnarray}
 where ${\bf \pi } =\frac{\hbar}{i} {\bf \nabla} - q {\bf A }$. 
 The remaining averaged correlators, involve products 
 $ G^R_{0 E} G^R_{0 E+\omega}$ and  $ G^A_{0 E} G^A_{0 E+\omega}$
 and are therfore by a factor $\hbar/(\tau E)$ smaller than the 
 conductivity, and can be disregarded for small disorder 
 $\hbar/\tau \ll  E $. 
In order to insert the Kubo-Greenwood formula in  the 
 saddle point expansion of the nonlinear sigma model, 
 it is convenient to rewrite the propagator in 
$F$ as
$$ G_{0 E} = \frac{1}{2} G^R_{0 E} ( 1 + \Lambda ) + 
\frac{1}{2} G^A_{0 E} ( 1 -  \Lambda). $$ 
 Then, we can use,  that  
$$Tr [ \sum_{\alpha = 1}^d \sum_{s = \pm} ( 1 + s \Lambda ) U ( \nabla_{\alpha}  \bar{U})
( 1 - s \Lambda ) U ( \nabla_{\alpha}  \bar{U})]
 = - Tr[ ( {\bf \nabla} Q)^2],$$. 
 Using the Kubo formula, Eq. (\ref{kubo}),  
  this functional of $Q$ simplifies to, 
\begin{eqnarray} \label{free2db}
F &=& \frac{h }{16 e^2}  
\int d {\bf x} \sum_{i=x,y}  \sigma (\omega=0)_{ii} ({\bf x})
 STr \left[ ( { \nabla_i} Q({\bf x} ))^2 \right] 
\nonumber \\ 
&-& \frac{1}{16} \frac{h}{e^2}   \int d {\bf x}  \sigma (\omega=0)_{xy}
({\bf x}) STr \left[ Q \partial_x Q \partial_y Q \right],
\end{eqnarray}
where $ \sigma (\omega=0)_{xy} =
 \sigma^I (\omega=0)_{xy}
({\bf x}) + \sigma^{II} (\omega=0)_{xy}$ 
 where $ \sigma^I (\omega=0)_{xy}$ is the dissipative part of the 
 Hall conductivity\cite{ando,pruisken,hajdu}.

\newpage 

\end{document}